\documentstyle[seceq,epsf]{ptptex}

\title{
Implications of the Discovery of a Millisecond Pulsar in SN 1987A 
}

\author{
Shigehiro {\sc Nagataki}$^{1}$
and Katsuhiko {\sc Sato}$^{1,2}$
}

\inst{
$^1$Department of Physics, School of Science, University of Tokyo\\
Tokyo 113-0033, Japan
\\
$^2$Research Center for the Early
Universe, School of Science, University of Tokyo\\
Tokyo 113-0033, Japan}

\recdate{
}

\abst{
From the observation of a millisecond pulsar in SN 1987A, the following
implications are obtained. 1) The pulsar spindown in SN 1987A is caused
by radiating gravitational waves rather than by magnetic dipole
radiation and/or relativistic pulsar winds. 2) A mildly deformed shock
wave would be formed at the core-collapse and explosion in SN 1987A,
which is consistent with the conclusion given in Nagataki (2000).
3) The gravitational waves from the pulsar should be detected in several
years using a Fabry-Perot-Michelson interferometer as the
gravitational detector, such as LIGO and TAMA.
4) The neutrino oscillation model is not a promising model for the
explanation of the kick velocity of the pulsar in SN 1987A.
The hydrodynamical instability model is more favored.
}

\begin{document}

\maketitle

\section{Introduction} \label{intro}
\indent

Since there is little informations concerning the pulsar in the
remnant of SN 1987A,
its properties, such as angular velocity of rotation,
strength of magnetic fields, and total baryon mass, have been treated as
free parameters or output
parameters.\cite{rf:kumagai93}$^{-}$\cite{rf:nagataki00}
However, Middleditch et al. (2000) reported the
discovery of an optical pulsar whose frequency is 467.5 Hz and spindown
rate is (2--3)$\times10^{-10}$ Hz s$^{-1}$.\cite{rf:mid00}
Since some free parameters appearing
in previous papers are constrained by this discovery, we consider
its implications in this paper. In section~\ref{mag}, we
show that the spindown is caused by radiating gravitational waves
rather than by magnetic dipole radiation and/or relativistic pulsar winds.
We also determine constraints on the strength of the magnetic field
of the pulsar.
In section~\ref{jet}, we discuss the effects of the
proto-neutron star's angular momentum on the dynamics of the
core collapse of the progenitor of SN 1987A. The
amplitude of the gravitational waves from the pulsar and its detectability
are discussed in section~\ref{grav}. Implications of the kick velocity
of the newly-born pulsar are presented in section~\ref{kick}.

\section{Origin of the pulsar spindown}
\label{mag}
\indent

Middleditch et al. (2000) reported that the spindowns
(2--3$\times 10^{-10}$ Hz s$^{-1}$) of the 2.14 ms pulsations should be
caused by radiating gravitational waves. This is because the relation
between the spindown rate and its modulation period can be explained
at the same time by adding the non-axisymmetric component of
the moment of inertia ($\delta I$) to a spherical neutron star whose
moment of inertia is $I$.
It is true that this conclusion is curious, because the spindown
of a normal pulsar is believed to be caused by magnetic dipole
radiation\cite{rf:pacini67} and/or relativistic pulsar
winds.\cite{rf:goldreich69} However, we want to emphasize that their
conclusion is supported by the recent UVOIR bolometric light curve.
We can easily calculate the decreasing rate of the rotational kinetic
energy of the pulsar as 
\begin{eqnarray}
    \label{eq:1}
\frac{dE}{dt} \equiv \dot{E}  =I \Omega \frac{d \Omega}{dt}, 
\end{eqnarray}  
where $\Omega$ is the angular velocity of the pulsar. Assuming that
the pulsar is spherical and has a constant density, the moment of inertia
of the pulsar can be expressed as
\begin{eqnarray}
    \label{eq:2}
I = 1.1 \times 10^{45} \left( \frac{M}{1.4 \; M_{\odot}} \right) 
\left( \frac{R}{10 \; \rm  km}  \right)^2  \;\;\; \rm [g \; cm^2].
\end{eqnarray} 
Thus, using the observation of $\Omega$ and $\dot{\Omega}$, we can
estimate $\dot{E}$ as
\begin{eqnarray}
    \label{eq:3}
\dot{E} = - (4-6) \times 10^{39} \left( \frac{M}{1.4 \; M_{\odot}} \right) 
\left( \frac{R}{10 \; \rm  km}  \right)^2  \;\;\; \rm [erg \; s^{-1}],
\end{eqnarray}
which is much larger than the UVOIR bolometric luminosisy, (1--2)
$\times 10^{36}$ erg $\rm s^{-1}$.\cite{rf:suntzeff99}
This discussion strongly supports that the pulsar spindown is caused
by radiating gravitational waves. Otherwise, the supernova remnant
would become much brighter.

We must also check whether the remnant is bright in other wavelengths,
such as radio, X-ray, and gamma-ray. If the brightness of the remnant
in these frequencies is not too large, we can confirm more strongly
our hypothesis that the pulsar spindown is caused by radiating
gravitational waves.

It is reported that the radio emission
spectrum is well fitted as\cite{rf:gaensler97}
\begin{eqnarray}
    \label{eq:4}
S \sim 10^{-15} \left(  \frac{\nu}{1 \; \rm GHz} \right)^{-1}
\;\;\; \rm [erg \; s^{-1} \; cm^{-2} \; GHz^{-1}].
\end{eqnarray}
In obtaining this expression, Gaesnler et al.\cite{rf:gaensler97} used data
at frequencies of
1.4, 2.4, 4.8, and 8.6 GHz. Therefore, when we assume that this
power-law fitting
holds at all radio frequencies, we can estimate its luminosity as
\begin{eqnarray}
    \label{eq:5}
L_{\rm radio} \sim 3 \times 10^{32} \left( \frac{D}{50 \; \rm kpc } \right)^2
 {\rm log_e}\left( \frac{\nu_{\rm max}}{\nu_{\rm min}}      \right),
\end{eqnarray}
where $D$ and $\nu$ are the distance from the Earth to the remnant and
the radio frequency, respectively. Assuming that the distance
is 50 kpc,\cite{rf:mccall93} we find that the luminosity in the
radio band is much smaller than the rate at which the rotational
energy decreases. In fact, unless $\nu_{\rm min}$ is as small as
$10^{-10^{6}}$ Hz,
the luminosity in the radio band is not comparable to the rate at which
the rotational energy decreases. Moreover, it is generally believed that the
radio emission does not come from the pulsar but from the synchrotron
emission of electrons that is generated when the shock encounters
circumstellar matter.\cite{rf:gaensler97}

An upper limit for X-rays of 2.3$\times 10^{34}$ erg $\rm s^{-1}$
(0.5 -- 2 keV) has been placed by the $Chandra$ observations of the
remnant.\cite{rf:burrows00} They also discussed that this low upper limit
is not surprising in view of calculations showing that debris
should still be opaque to soft X-rays.\cite{rf:fransson87}
Hence we can conclude that the rapid rate at which the rotational
energy decreases cannot be explained by the emission of the soft X-rays.

The remnant is thought to be
transparent to hard X-rays and gamma-rays.
In fact, Fransson and Chevalier\cite{rf:fransson87} reported
that the energy corresponding to unity of the absorption optical depth of
the ejecta can be well represented by the formula
\begin{eqnarray}
    \label{eq:6}
E(\tau = 1) = 81 \left( \frac{M_c}{1 \; M_{\odot}} \right)^{0.36} 
\left( \frac{V_c}{2500 \; {\rm km \; s^{-1} }} \right)^{-0.72}
\left( \frac{t}{1 \; {\rm yr }} \right)^{-0.72} \;\;\; \rm  [{\rm keV}], 
\end{eqnarray}
where $M_c$, $V_c$, and $t$ are the mass inside the O/He interface, 
the expansion velocity of the core, and the time from the explosion,
respectively. When we adopt $M_c$ = 3.7$M_{\odot}$,\cite{rf:hashimoto95}
$V_c$ = 2500 km $\rm s^{-1}$,\cite{rf:nagataki00} and $t$ = 5 yr,
we obtain
$E(\tau =1)$ = 40 keV. Thus the situation here is different from that for
soft
X-rays; that is, the remnant is thought to be transparent to hard
X-rays and gamma-rays. As for the data at these frequencies, the
upper limit of the spectrum is rather rough, and published data
are not new. The gamma-ray continuum
from April 4, 1989 can be fit as\cite{rf:palmer93}
\begin{eqnarray}
    \label{eq:7}
\frac{dN}{dE} = 1.6 \times 10^{-5} \left(   \frac{E}{100 \; \rm keV}
\right)^{-1}
\;\;\; \rm [photons \; cm^{-2} \; s^{-1} \; keV^{-1}]
\end{eqnarray}
for the energy range 50-800 keV. The total energy flux can be obtained
as
\begin{eqnarray}
    \label{eq:8}
L_{\rm gamma} \sim 8 \times 10^{37} \left(   \frac{D}{\rm 50 \; kpc}  \right)^2
\left( \frac{E_{\rm max}}{\rm 100 \; keV}   \right),
\end{eqnarray}
where $E_{\rm max}$ is the maximum energy of the gamma-ray photons.
It is generally believed that these gamma-rays come from
radioactive nuclei such as $\rm ^{56}Co$ and $\rm ^{57}Co$.\cite{rf:kumagai89}
Moreover, the Crab nebula, whose energy source is the central pulsar,
is brightest in the X-ray band. Thus it is difficult to think that the
remnant in SN 1987A could be brightest in the gamma-ray band and its luminosity
could be comparable to the rate at which the rotational energy decreases.
In the case that the total gamma-ray luminosity is found by future
observations to be as large as
the rate at which the rotational energy decreases,
we will have to consider the serious problem of producing gamma-rays only.
That is, we have to face the difficult problem of determining a
mechanism that produces only gamma-rays and no photons in other
energy bands from the pulsar. 

For the reasons mentioned above, we think that it is difficult to
explain the observed spindown
with the magnetic dipole model and/or relativistic pulsar wind model.
However, it is necessary to fix a strict upper limit of the present
gamma-ray flux from SN 1987A in order to conclude with certainty
that the pulsar spindown is not caused
by radiating photons and/or ejecting relativistic particles but by
gravitational waves.

We can also give un upper limit for the strength of the
magnetic field of the pulsar.
From the magnetic dipole model and/or the relativistic pulsar wind model,
the rate at which the energy decreases can be written\cite{rf:goldreich69}
\begin{eqnarray}
    \label{eq:9}
\dot{E} = - \frac{B_p^2 R^6 \Omega^4}{c^3},
\end{eqnarray}
where $B_p$ is the strength of the magnetic field at the magnetic
pole of the star. From Eqs.~(\ref{eq:3}) and~(\ref{eq:9}), we can
derive the upper limit for $B_p$ as
\begin{eqnarray}
    \label{eq:10}
B_p \le (4 - 5) \times 10^{10} \left( \frac{M}{1.4 \; M_{\odot}}    \right)
\left( \frac{10 \; \rm km}{R}     \right)^4   \;\;\;  \rm [G].
\end{eqnarray} 
Since the strength of the magnetic field is so weak, we believe that a
neutron star
without hot spots will be found in X-ray bands in the near future.
The lack of hot spots implies that
the surface temperature of such a neutron star will be approximately
uniform. Such a neutron star should be found when the optical depth
becomes sufficiently low.
We also add a comment on the weakness of the magnetic field of the
pulsar in SN 1987A. Since it is apparently weaker than the typical
one, it seems to be suggested that the strength of the magnetic field
of a newly born pulsar evolves as a function of time.
Thus we may be able to observe in the
near future the magnetic field of a pulsar growing stronger
when we can observe the pulsar activity directly.

Finally, we consider the possibility of radiating gravitational waves
to a sufficient extent to explain the observed pulsar spindown.
Middleditch et al. (2000) concluded that the required non-axisymmetric
oblateness ($\epsilon$) is $\sim 10^{-6}$, since a slightly deformed,
homogeneous ellipsoidal pulsar with moment of inertia $I$ and
ellipticity $\epsilon$ radiates energy in the form of gravitational waves
at a rate
\begin{eqnarray}
    \label{eq:11}
\dot{E}_{\rm GW} = - \frac{32}{5}\frac{G}{c^5}I^2  \epsilon^2 \Omega^6.
\end{eqnarray}
Here $\epsilon$ is defined as
\begin{eqnarray}
    \label{eq:12}
\epsilon = \frac{a - b}{ (a+ b)/2},
\end{eqnarray}
where $a$ and $b$ are the equatorial semiaxes.

We should discuss where and why a non-axisymmetric component of
the moment of inertia is realized in a neutron star.
Here we have to note that the average density of a pulsar is about
5$\times 10^{14}$ g cm$^{-3}$. Thus it is meaningless to consider a
'mountain' on the surface of a neutron star, where the density is
about $10^{9}$ g cm$^{-3}$ and the density scale height is only
$\sim$ 1cm.\cite{rf:iida97} This is because the contribution of the
mountain on the surface of the neutron star is too little to explain
the non-axisymmetric component of the moment of inertia.
Rather, we should consider
density fluctuation in the inner crust, where the typical density
is sufficiently high and the contribution to the moment of inertia is
not negligible. In particular, Lorenz et al.\cite{rf:lorenz93}
reported that there may be a nuclear `pasta' at the innermost
region of the inner crust.\cite{rf:watanabe00} 
In this nuclear pasta region,
we can easily guess that non-uniform crystallization due to
the rapid cooling of the newly born neutron star will result in such
a non-axisymmetric component of the moment of inertia.
It is a very important
task to estimate the nucleation rate and the growth rate of such
a crystal
in the pasta region. This should give information on the
non-axisymmetric component of the moment of inertia in neutron stars.

\section{Implications for the jet-like explosion in SN 1987A}\label{jet}
\indent

Effects of rotation on the dynamics of collapse-driven supernovae
have been investigated in many works.\cite{rf:yamada94}$^,$
\cite{rf:nagataki00}$^,$\cite{rf:shimizu94}$^,$\cite{rf:muller80}
However, the initial rotational energy has been given parametrically
in these works, because we have little information on it.
Because we now have information on the rotational energy of the newly
born pulsar in SN 1987A, we can now carry out further analysis.
First, we estimate the initial period of the pulsation.
From Eqs.~(\ref{eq:1}) and~(\ref{eq:11}), the initial period can be
estimated as
\begin{eqnarray}
    \label{eq:13}
T_{\rm i} = T_{\rm o} \left(  1 - 4 \frac{t}{T_{\rm GW}}      \right),
\end{eqnarray}
where $T_{\rm o}$, $t$, and $T_{\rm GW}$ are the present
period ($\sim 2.14$ ms),
present time $\sim 5$ yr, and -$\Omega_{\rm o}/\dot{\Omega}_{\rm o}$,
respectively. From Eq.~(\ref{eq:13}), the initial period of pulsation
can be estimated as (1.9--2.0)ms. Even if we assume that the pulsar spindown
is caused by magnetic dipole radiation and/or relativistic pulsar
winds, the estimated initial period changes little.
Now we can estimate the ratio of the rotational energy relative to
the gravitational binding energy at the moment of the core-collapse  
($T$/$| W |_{\rm init}$). It is estimated as
\begin{eqnarray}
    \label{eq:14}
T/ | W |_{\rm init} && = \frac{25G}{12 c^2}q^2 \left(  \frac{M}{R} \right) \\
&& \sim 4.3 \times 10^{-3} \left(     \frac{M}{1.4M_{\odot}} \right)
\left(  \frac{1000 \; {\rm km}}{R}     \right) q^2
\end{eqnarray}
where $q$ = $Jc/GM^2$ = $I \Omega c/GM^2$ is the dimensionless angular
momentum. Since the value for $q$ can be estimated as
\begin{eqnarray}
    \label{eq:15}
q = 0.2 \left(  \frac{1.4M_{\odot}}{M}  \right)
\left(  \frac{R}{10 \; \rm km}  \right)^2
\left(  \frac{2 \; \rm ms}{P}  \right),
\end{eqnarray}
$T$/$| W |_{\rm init}$ for the progenitor of SN 1987A can be estimated
to be 1.7 $\times 10^{-4}$.

We wish to stress the fact that this estimated value is smaller than
the values assumed in the study of Yamada and Sato (1994),
in which an extremely deformed shock wave is formed. Thus, it can be
easily guessed that a mildly deformed shock wave was formed in the
core of SN 1987A. This is consistent with the conclusion reached
in Nagataki (2000).

We can also estimate the ellipticity ($e$) of the proto-neutron star in
SN 1987A from the rotational energy. Since the relation between $e$ and
$T$/$| W |$ can be written as\cite{rf:shapiro83}
\begin{eqnarray}
    \label{eq:16}
\frac{T}{|W|} = \frac{3}{2e^2}\left( 1 - \frac{e(1-e^2)^{1/2}}{\sin ^{-1} e}
       \right)           -1,
\end{eqnarray}
$e$ can be estimated as $\sim$ 0.25. Here we have assumed that the mass and
radius of the proto-neutron star are 1.4$M_{\odot}$ and 20 km, respectively.
This means that the ratio of the semimajor axis relative to the semiminor
axis of the proto-neutron star is $\sim$ 1.03.
It should be noted that this value is smaller than that assumed in the
work of Shimizu et al. (1994), in which an extremely
deformed shock wave is formed due to the effects of asymmetric neutrino
heating from the deformed neutrinosphere. 
This discussion given here also supports the conclusion reached
in Nagataki (2000) that a mildly deformed shock wave is
required in order to realize the appropriate matter mixing and
explosive nucleosynthesis in SN 1987A.

It is a very important task to perform numerical simulations
in which the effects of rotation and neutrino heating are included
in order to make an appropriate model for SN 1987A in which a
mildly deformed shock wave and a rotating neutron star with a
period of 2 ms are formed. Such a model will help us to understand
more clearly the system SN 1987A and the roles of rotation and asymmetric
neutrino heating on the dynamics of collapse-driven supernovae.

\section{Gravitational waves from the pulsar}\label{grav}
\indent

We can estimate the amplitude of the gravitational waves from the
pulsar in SN 1987A. The energy release rate due to the gravitational
waves can be written
\begin{eqnarray}
    \label{eq:17}
\dot{E} = \frac{c^3}{16 \pi G} \Omega^2 \langle  h \rangle^2 \times 4\pi D^2,
\end{eqnarray}
where $\langle h \rangle$ is the average dimensionless amplitude
of the gravitational waves
at the distance $D$ from the pulsar. From Eqs.~(\ref{eq:11})
and~(\ref{eq:17}), $\langle h \rangle$ can be estimated as
\begin{eqnarray}
    \label{eq:18}
\langle h \rangle && \sim 5.1 \frac{G}{c^4 D}I \epsilon \Omega^2 \\
&& \sim 4.7 \times 10^{-26} 
\left( \frac{I}{1.1\times 10^{45} {\; \rm g \; cm^2}}   \right)
\left( \frac{\epsilon}{10^{-6}}      \right)
\left( \frac{\Omega}{2936 \; \rm rad \; s^{-1}}   \right)^2
\left( \frac{50 \; \rm kpc}{D}         \right).
\end{eqnarray}
Thus, the required time to detect the gravitational wave from the
pulsar using the Fabry-Perot-Michelson interferometer as the
gravitational detector is
\begin{eqnarray}
    \label{eq:19}
\Delta T = 4.2 \left( \frac{h'}{3 \times 10^{-22} [{\rm 1/\sqrt{Hz}}]}
\right)^2   \;\;\; \rm [yr],
\end{eqnarray}
where $h'$ is the sensitivity of the detector at 2$\times$467.5 = 935 Hz,
which has units of $1/\sqrt{\rm Hz}$. We can find that detection of
the gravitational wave from the pulsar is possible within a reasonable
time when gravitational detectors such as LIGO\cite{rf:abramovici92}
and TAMA\cite{rf:tsubono96}, whose sensitivities are of
order $h'$$\sim$$10^{-22}$ $\rm Hz^{-1/2}$, are running.

\section{Implications on the kick velocity of a newly-born pulsar}
\label{kick}
\indent

It is a well known fact that pulsars in our galaxy have velocities
much greater than those of ordinary stars.\cite{rf:hla93}
It is reported that their
transverse speeds range from 0 to $\sim$ 1500 km $\rm s^{-1}$ and
their mean three-dimensional speeds are $\sim$450 $\pm$ 90 km $\rm
s^{-1}$.\cite{rf:ll94}
There are many theoretical models to explain the
pulsar kick. According to one, a neutron star in a binary system can escape
from the system with rapid speed due to a supernova explosion of the
nascent star.\cite{rf:ggo70}
There are also many models in which
effects of asymmetric supernova explosions are taken into
consideration.\cite{rf:bh96}\cite{rf:ks96} For example, it is reported
that neutrino oscillations, biased by the magnetic field, alter
the shape of the neutrino sphere in a cooling proto-neutron star and
are the origin of the kick velocity of the pulsar.\cite{rf:ks96}
In another work, Burrows and Hayes (1996) pointed out the possibility
that hydrodynamical instabilities may be the origin of the pulsar
kick. However, there are too few
observations to determine which model is the most promising one.

It is suggested that the pulsar in SN 1987A also has a kick velocity
and is moving toward the south region of the
remnant.\cite{rf:fargion95}$^,$\cite{rf:fargion95b}$^,$\cite{rf:nagataki00}$^,$\cite{rf:nagataki00b}
As discussed in section~\ref{mag}, it is suggested that the strength of
magnetic fields on the surface of
the pulsar in SN 1987A is very weak. Therefore it is concluded that
a neutrino oscillation model like that of Kusenko and
Segr$\rm \grave{e}$ (1996),
which requires magnetic fields of order $10^{14}$ G, is not
promising. This is the first investigation with the purpose of selecting the
best model of the kick velocity using observational data.
We will be able to give further discussion when
we obtain more precise data on the pulsar in SN 1987A.
We hope there will be further observations of this pulsar at
many frequencies of
photon and gravitational waves in the near future so that we can continue
our investigation and confirm the report of the discovery
presented by Middleditch et al. (2000).

\section*{Acknowledgments}

We would like to thank K. Tsubono, K. Iida and S. Yoshida for
valuable information and useful discussions.
This research has been supported in part by a Grant-in-Aid for
Center-of-Excellence (COE) Research (07CE2002) and by the Scientific
Research Fund (199908802) of the Ministry of Education, Science, Sports and
Culture of Japan and by the Japan Society for the Promotion of Science
Postdoctoral Fellowships for Research Abroad.  


\end{document}